\documentclass[preprint, 12pt]{elsarticle}
\usepackage{soul}
\soulregister\ref7
\soulregister\cite7

\usepackage{tabularx}
\usepackage{ifpdf}
\usepackage{graphicx}
\usepackage{multirow}
\usepackage{natbib}
\usepackage{amsmath,amsfonts,amssymb}
\usepackage{algorithmic}
\usepackage{textcomp}
\usepackage{array}
\usepackage{threeparttable}
\usepackage{stfloats}
\usepackage{url}
\usepackage{tabu}
\usepackage{todonotes}
\usepackage[bottom]{footmisc}
\usepackage{hyperref}
\usepackage{subfig}
\usepackage[normalem]{ulem}
\usepackage{cleveref}

\usepackage{booktabs}
\newcommand{\ra}[1]{\renewcommand{\arraystretch}{#1}}

\usepackage{tikz}
\usetikzlibrary{shapes, arrows, positioning}

\tikzstyle{startstop} = [rectangle, rounded corners, minimum width=1.5cm, minimum height=1cm,text centered, draw=black]
\tikzstyle{iter} = [rectangle, minimum width=1.5cm, minimum height=1cm,text centered, draw=black]
\tikzstyle{io} = [trapezium, trapezium left angle=70, trapezium right angle=110, minimum width=3cm, minimum height=1cm, text centered, text width=3cm, draw=black]
\tikzstyle{process} = [rectangle, minimum width=5cm, minimum height=1cm, text centered, text width=9cm, draw=black]
\tikzstyle{decision} = [diamond, aspect=2, minimum width=1cm, minimum height=1cm, text centered, draw=black]
\tikzstyle{circ} = [circle, minimum size = .75cm, centered, draw=black]
\tikzstyle{elip} = [ellipse, text width=2.2cm, minimum width = 1.25cm, minimum height = 1.9cm, text centered, draw = black]
\tikzstyle{line} = [draw, -latex']
\tikzstyle{arrow} = [thick,->,>=stealth]

\begin{document}
\begin{frontmatter}
%
\title{A Coupled Karhunen--Lo\`{e}ve and Anisotropic Sparse Grid Interpolation Method for the Probabilistic Load Flow Problem}
%
%
%

\author{Brandon~Johnson,~Nathan~L.~Gibson,~and~Eduardo~Cotilla-Sanchez%
        }
\address{Oregon State University, Corvallis, OR, 97330, USA}
\begin{abstract}
In the traditional load flow analysis, a key assumption is that the input variables, i.e., generator output and customer demand, are fixed in time and the associated response has no variability. This assumption, however, is no longer valid as the adoption of renewable energy resources add more variability and uncertainty to the modern electrical system. Addressing these concerns is the definition of the Probabilistic Load Flow (PLF) problem. The challenge of the PLF problem lies in handling high-dimensional input uncertainties and the non-linearity of the load flow equations. The most straightforward way to address these problems, but at the cost of computational time, is to perform a Monte Carlo method. This work, however, solves these problems---accuracy, high-dimensionality, and computational time---with a coupled Karhunen-Lo\`{e}ve (KL) expansion and Anisotropic Sparse Grid algorithm. The proposed method is implemented and tested on the IEEE 118-bus test system and a modernized version of the IEEE Reliability Test System--1996, the Reliability Test System--Grid Modernization Lab Consortium (RTS-GMLC). Results for the 194-dimensional case show a decrease in computational time when compared to the 10,000 sample Monte Carlo method given a bound on mean and standard deviation error.

\end{abstract}


\end{frontmatter}

\section{Introduction}
Load flow studies are one of the most essential tools for grid operators because they provide a clear picture of the operating conditions of the power system. Traditionally, the solution to a load flow analysis is assumed to be deterministic; that is, the system variables contain no randomness. In recent years, distributed energy resources, such as wind and solar, are being adopted at an increasing rate causing the power system to be subjected to increasing levels of variability. The intrinsic variability of renewable resources stems from the inability to accurately predict local weather patterns at relatively small and discrete time-steps. As a result of these changes in the power system, it is no longer sufficient to assume that both generation and demand are constants in the load flow problem. The reality is that it is much more accurate to assess the power systems response to a range of possible inputs. 

The Probabilistic Load Flow (PLF) problem was first proposed by Borko\-wska \cite{Borkowska1974} with the goal of handling uncertainties, e.g., variability in the energy supply, within a power flow analysis. Since then, there are two major ways of solving the PLF problem:~numerically, i.e., using a Monte Carlo method \cite{Morales2007, Chen2008}, or analytically, i.e., using a convolutional method \cite{Allan1981}. As in the study of other complex systems, in power systems Monte Carlo methods allow for a simplistic approach for solving the PLF, but come with the downside of high computational burden. As a result of this characteristic, Monte Carlo methods are more suited as a baseline for checking the accuracy of other methods. Analytical methods, on the other hand, are mathematically complex and are subject to inaccuracies that vary depending on assumptions and approximations, but have the capability to solve the PLF problem in real time--as it is needed in operations and control. 

The objective of analytical methods is to obtain the probability distribution function (PDF) and cumulative distribution function (CDF) of the state vector, i.e., active and reactive power injections, voltage, and angle. The difficulty of this problem lies within the non-linear nature of the load-flow equations and that the uncertain parameters that affect the power system are unlikely to be independent. To remedy these two problems researchers have employed these two relatively simple steps:~1) linearize the load flow equations and 2)~assume that the uncertain parameters are independent; these two steps alone enable the use of convolutional techniques. Unfortunately, convolutional methods have three major downsides:~1) linearizing the power-flow equations around the operating point is sufficient only when deviations are close to the operating point 2) in most cases, the assumption of statistical independence of input variables is often false \cite{Kloubert2017} and 3) convolutional techniques require a large amount of memory. For example, in the case where two random variables are represented by discrete functions with $k$ impulses, the resulting convolution will have $k$ times $k$ impulses \cite{Zhang2004}. 

Recent work in this area has attempted to forego the convolutional metho\-d and move in favor of methods that seek to approximate the target output variables using cumulant and polynomial expansion methods. References \cite{Zhang2004, Fan2012} use the properties of cumulants--an alternative to moments--and the Gram-Charlier expansion to reduce the computation time while still maintaining a comparable accuracy to that of Monte Carlo methods. References \cite{Appino2017, Wu2017} have used a polynomial chaos expansion to reformulate the stochastic problem into a purely deterministic one without the use of random sampling. Similarly, Sun \textit{et. al} \cite{Sun2018} applied a sparse polynomial chaos expansion to a small test system with four wind farms and 21 random load sources. Reference \cite{Zhou2017} uses perturbation methods, derived from Galerkin methods, to resolve issues using polynomial approximations of the load flow equations. While the main benefit of these methods are the significant reduction in computation time, the assumptions made to calculate the output PDF lead to inaccuracies. 

A third group of methods, called approximation methods, seek to leverage the efficiency of deterministic solvers to build the output PDFs, much like Monte Carlo methods, but with far fewer simulations. The first method of this kind--introduced in 1975 by Rosenblueth--is called the Point Estimate Method (PEM) \cite{Rosenblueth1981}. The PEM bundles statistical information using the first few central moments of the input variable into $K$ points for each variable, called \textit{concentrations}. Rosenblueth's original work showed that the probabilistic problem can be reduced to $2^m$ deterministic simulations, where $m$ is the number of random input variables. Since then, adaptations have been made to the original problem that can reduce the number of simulations to as little as $2m + 1$ while still retaining a high-level of accuracy \cite{Harr1989, Li1992, Hong1998}. While the PEMs possess the speed of the analytical solutions and the accuracy of Monte Carlo methods, they can only produce statistical moment information of the output variables. One solution to overcome this problem is to employ Stochastic Collocation (SC) methods.

While technically an approximate method, SC methods are able to produce the necessary PDFs and CDFs to assess the risk of a system with uncertain input variables. Much like PEMs, one creates a set of nodes, called \textit{collocation} points, and then the model is solved using a deterministic solver at each node. One approach to the SC problem is to use the interpolation approach with Tensor Grids \cite{Xiu2009, Dongbin2010}. Given the high-dimensionality of the PLF problem, it is more practical to use a subset of full Tensor Grids called Smolyak sparse grids \cite{Smolyak1963}. Sparse grids are capable of reducing the impact of high-dimensionality that afflicts other approximation methods. The authors in \cite{Tang2016, Ni2017} use a dimension-adaptive sparse grid interpolation, in conjunction with Copula Theory principles, to explore the impact of dependent random variables on the PLF problem. While being fairly accurate, this method is computationally burdensome due to having to determine the sparse grid as the problem evolves. This rules out the option of parallelizing the deterministic load flow simulations due to the interdependence of each simulation.

We should note that sparse grids alone do not reduce the dimension of the problem, therefore, we propose to apply the Karhunen--Lo\`{e}ve (KL) expansion \cite{Kosambi1943, Karhunen1947, Loeve1978} to the random fields. The KL-expansion is a dimension reduction technique that, much like Principal Component Analysis (PCA), seeks to represent a random process as a series combination of a complete set of deterministic functions with corresponding random coefficients. Our work also proposes coupling the KL-expansion with an \textit{anisotropic} sparse grid \cite{Nobile2008}, a more generalized Smolyak sparse grid, further improving the performance of approximate methods. By reducing the number of dimensions and assigning importance to each dimension, our method provides a significant computational savings while maintaining a comparable accuracy to existing methods. Furthermore, the PDF and CDF of the output can be computed directly by using the sparse grid interpolant.

The remainder of the paper is organized as follows. Section~\ref{sec:KL} includes a mathematical formulation for the KL-expansion. Section~\ref{sec:SC} provides a formal background for SC and the mathematical framework for anisotropic sparse grids. Then, in Section~\ref{sec:KL+ASG} we discuss how both the KL-expansion and ansiotropic sparse grids are coupled to solve the PLF problem. Thereafter, in Section~\ref{sec:setup} we outline the implementation and metrics used for accuracy analysis. Lastly, Section~\ref{sec:results} shows the results of implementing the proposed method and the concluding remarks follow in Section~\ref{sec:conclusion}.

\section{Dimension Reduction and the Karhunen--Lo\`{e}ve (KL) Expansion}\label{sec:KL}

\subsection{Mathematical Formulation}
We begin by assuming there is a (spatial) correlation among the random input variables, and thus treat each as a sample of an underlying spatial stochastic process, which we represent efficiently using a KL-expansion, i.e., a series expansion in orthogonal eigenfunctions of the covariance structure. For example, multiple independent solar farms (in a given region) can be treated as a random process. One can treat them as either temporally (start producing near the same time) or spatially (similar solar irradiance, cloud coverage, etc.) correlated where the actual value of solar production is a random variable based on several other random variables. Consider the random process with two variables $Y(x, \omega)$ where $x$ is an index parameter (representing either time or space) and the outcome $\omega \in \Omega$, where $\Omega$ is the space of the underlying random variables. The random process $Y(x, \omega)$ has mean $\Bar{Y}(x)$ and covariance function $C(x, s) = \mathrm{Cov}(Y(x), Y(s))$. Then, $Y(x, \omega)$ has the orthogonal decomposition

\begin{equation}
    Y(x, \omega) = \Bar{Y}(x) + \sum_{n=1}^{\infty}\sqrt{\lambda_n}\phi_n(x)\xi_n(\omega),
    \label{eqn:KL}
\end{equation}

\noindent where $\{(\lambda_n, \phi_n)\}^{\infty}_{n=1}$ are the eigenvalues and orthogonal eigenfunctions, which are the solutions to a Frendholm integral equation of the second kind, defined as

\begin{equation}
    \lambda\phi(s) = \int C(x, s) \phi(x)dx.
    \label{eqn:frendholm}
\end{equation}

The random variables $\{\xi_n \}^{\infty}_{n=1}$ in (\ref{eqn:KL}) are uncorrelated and satisfy
\begin{align}
    E[\xi_n] &= 0, \quad \mathrm{Cov}[\xi_n, \xi_k] = \delta_{n k}, \quad n,k \geq 1,
    \label{eqn:frendholm_const}
\end{align}
where $\delta_{n k}$ is the Kronecker-delta function. It is often practical to truncate the KL-expansion in (\ref{eqn:KL}) to a finite number of terms $d$. The truncated KL-expansion then becomes

\begin{equation}
    Y(x, \omega) \approx \Bar{Y}(x) + \sum_{n=1}^{d}\sqrt{\lambda_n}\phi_n(x)\xi_n(\omega), \quad d \geq 1.
    \label{eqn:truncKL}
\end{equation}

The number of $d$ terms to keep in the truncated KL-expansion is often determined by examining the decay rate of the eigenvalues $\lambda_n$ as the index $n$ increases. The decay rate of the eigenvalues depends inversely on the correlation length of the random process \cite{Dongbin2010}. This implies that having a long correlation length (i.e.,~strongly correlated variables) results in a fast decay of the eigenvalues. Conversely, an uncorrelated process with zero correlation length proves to be the limit of this method in which there is no eigenvalue decay. It is common to select $d$ such that the remaining eigenvalue contributions are negligible. In practice, it is often possible to capture 90\% of the input variability with only 10 variables. 

\subsection{The Need for Stochastic Collocation}
While we will cover stochastic collocation in more detail in Section~\ref{sec:SC}, it is important to discuss why the KL-expansion is necessary in the first place. Stochastic collocation methods, like PEMs, are very desirable because they help solve the PLF problem non-intrusively. That is, one can leverage fast deterministic solvers on a given set of samples or collocation points. In this way, PEMs may also be considered a collocation method. Unfortunately, in large problem spaces collocation methods suffer from something called the ``curse of dimensionality'', i.e., the problem scales poorly in high dimensions. One way to overcome this limitation is to use a sparse set of collocation points. Although this lessens the impact caused by high-dimensionality, it does not effectively reduce the number of dimensions in the original problem space. It is in this way how--by using the KL-expansion to reduce the dimensionality--collocation methods become computationally tractable. We choose each $\xi_n$ in (\ref{eqn:truncKL}) to be an independent random variable uniformly distributed over the interval $[-1, 1]$. This then allows us to use quadrature rules (please see Section~\ref{sec:KL+ASG}) to deterministically sample each $\xi_n$; thus, mapping the stochastic problem into a deterministic one. Note that by choosing $\xi_n$ to be uniformly distributed over $[-1, 1]$ means one must scale the KL-expansion by $\sqrt{3}$ to ensure that each variable has a variance of 1.

\section{Stochastic Collocation and Tensor Grid Interpolants}\label{sec:SC}
The goal of SC methods is to solve a set of governing equations (e.g.,~a set of partial differential equations) at discrete nodes called collocation points. In this context, Monte Carlo methods can also be considered a collocation method where the collocation points are determined randomly. In cases where the input distributions are compatible with Askey scheme orthogonal polynomials, it is more efficient to use cubature rules to determine the nodal set \cite{Dongbin2010}. With this in mind, the following section serves as the mathematical foundation for determining the nodal set used for tensor grid and sparse grid collocation methods. For more details on the methods presented here please see references \cite{Smolyak1963, Griebel1998a, Gerstner2003, Nobile2008, Dongbin2010}.

\subsection{Tensor Product Collocation}

\begin{figure*}[!t]
    \centering
    \includegraphics[trim = {0.1cm .8cm 0.1cm .9cm}, clip, width = 1\linewidth]{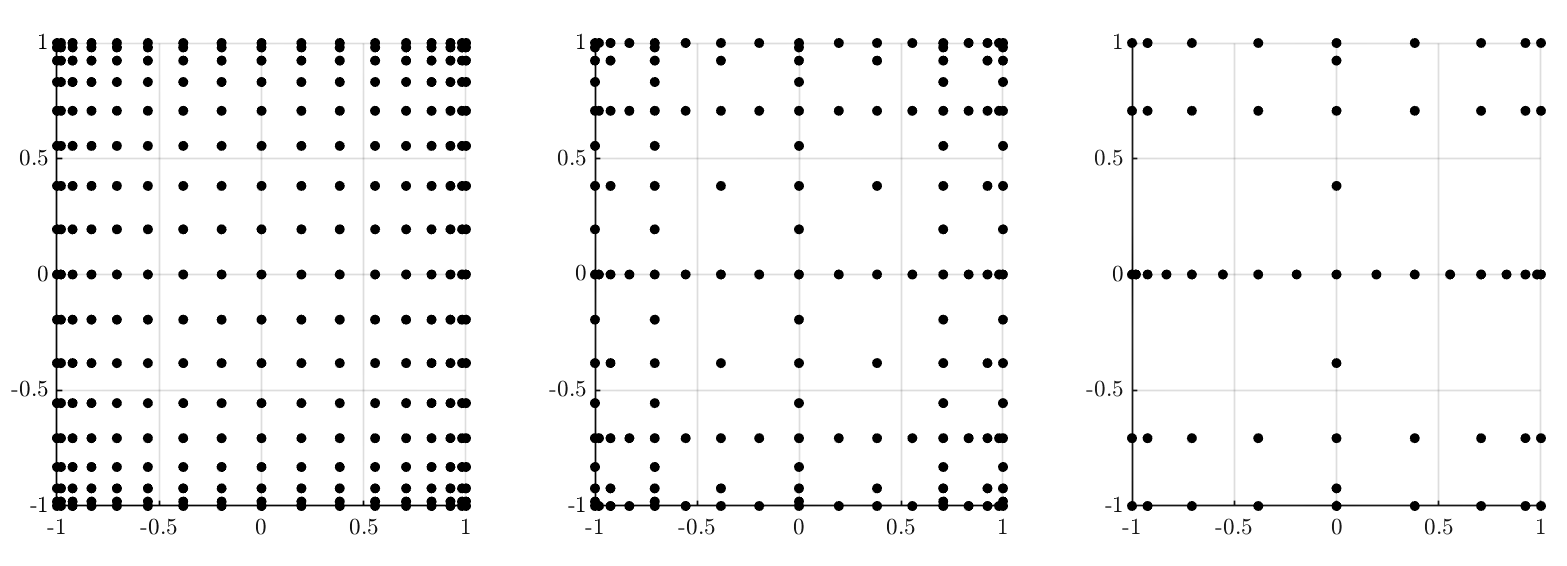}
    \caption{Two-dimensional nodes for one-dimensional extrema of the Chebyshev polynomials at level $k = 5$. Left:~Full Tensor Grid. The total number of nodes is 289. Middle:~Smolyak Sparse Grid. The total number of nodes is 161. Right:~Anisotropic Sparse Grid with $\gamma_2/\gamma_1 = 2$. The total number of nodes is 57.}
    \label{fig:grid_plots}
\end{figure*}

For multivariate cases with dimensionality $d > 1$, for any $1 \leq i \leq d$, let $Q_{m_i}$ be an interpolating operator such that
\begin{equation}
    Q_{m_i}[f] = \Pi_{m_i}f(Z_i) \in \mathbb{P}_{m_i}(Z_i),
    \label{eqn:inter_operator}
\end{equation}
is an interpolating polynomial of degree $m_i$, for a given function $f$ in the $Z_i$, variable by using $m_i + 1$ distinct nodes in the set $ {\Theta^{m_i}}_1 = \{{Z^{(1)}}_1,\ldots,{Z^{m_i}}_i\}$. Then the most straightforward approach to interpolating $f$ in the entire space $I_z \subset \mathbb{R}^d$ is to use the tensor product approach. That is,
\begin{equation}
    Q_M = Q_{m_1} \otimes \cdots \otimes Q_{m_d},
    \label{eqn:tensor_product}
\end{equation}
and the nodal set is
\begin{equation}
    \Theta_M = \Theta^{m_1}_1 \times \cdots \times \Theta^{m_d}_1,
    \label{eqn:tensor_nodal_set}
\end{equation}
where the total number of nodes is $M = m_1 \times \cdots \times m_d$.
In the case where the total number of points $M = m^d$, the interpolation error follows
\begin{equation}
    (I - Q_M)f[m] \propto M^{-\alpha/d},
    \label{eqn:inter_error}
\end{equation}
where the constant $ \alpha > 0$ depends on the smoothness of the function $f$ \cite{Dongbin2010}. For large dimension $d \gg 1$, the total number of points $M = m^d$ grows very fast for large $d$. Again, this is commonly referred to as the ``curse of dimensionality.''

\subsection{Sparse Grid Collocation}
A much more practical approach rather than the full Tensor Product Collocation are Smolyak sparse grids \cite{Smolyak1963}. The Smolyak sparse grid construction is based on the tensor product construction but they only represent a subset of the full tensor grid. The construction of the grid takes the form 

\begin{equation}
    Q_N = \sum_{N-d+1 \leq |\boldsymbol{i}| \leq N} (-1)^{N- |\boldsymbol{i}|}\binom{d - 1}{N - |\boldsymbol{i}|} \cdot (Q_{i_1} \otimes \cdots \otimes Q_{i_d}),
\end{equation}
where $N \geq d$ is an integer denoting the \textit{level} of the construction. The nodal set, i.e.,~the sparse grid, is

\begin{equation}
    \Theta_M = \bigcup_{N-d+1 \leq |\boldsymbol{i}| \leq N} (\Theta_{1}^{i_1} \times \cdots \times \Theta_{1}^{i_d}).
\end{equation}

It is typically desired for the grid to take a nested form. One popular choice for a nested grid are the Clenshaw-Curtis nodes, which are the extrema of the Chebyshev polynomials, and are defined as
\begin{equation}
    Z_{j}^{(i)} = -\cos{\frac{\pi(j-1)}{m_{i}^{k} - 1}}, \quad \quad j = 1 ,..., m_{i}^{k},
\end{equation}
where an additional index is introduced via the subscript $k$, often described as the level of the Clenshaw-Curtis grid (a higher level creates a finer grid). While there is not a closed form expression for the total collocation points $M$ in terms of $k$ and $d$, in high-dimensional problems the total number of points is estimated by

\begin{equation}
    M \sim 2^k d^k /k! \quad \quad d \gg 1.
\end{equation}

Although the curse of dimensionality has been lessened through the use of sparse grids, it still exists.

\subsection{Anisotropic Sparse Grid Collocation}
Until now the focus of sparse grids has been on methods that treat each dimension \textit{isotropically}; that is, each dimension is treated equally. This assumption, however, is only valid for problems where there is either a weak dimensional dependence or that dependence is unknown. If, on the other hand, the problem exhibits a strong dimension-dependent variation then it is desirable to use an anisotropic formulation of the sparse grid problem \cite{Nobile2008, Burkardt2012}.

Let $\mathbf{i}$ be a $d$-dimensional level vector where each dimension $n$ is defined by $i_n \in \mathbb{N}_+$ and let $\mathbf{\gamma} = \{\gamma_1, \gamma_2, \dots, \gamma_d\} \in \mathbb{R}^{d}_{+}$ be a $d$-dimensional weight vector for each stochastic dimension $\omega$. We then denote the anisotropic sparse grid of 0-based index $w$, spatial dimension $d$, and anisotropy vector $\gamma$, by $\mathbb{A}(w, d, \gamma)$, which has the form

\begin{equation}
    \mathbb{A}(w, d, \gamma) = \sum_{\mathbf{i} \in Y_{\gamma}(w, d)} c_{\gamma}(\mathbf{i})(Q_{i_1} \otimes \cdots \otimes Q_{i_d}),
    \label{eqn:aniso}
\end{equation}
with
\begin{equation}
    c_{\gamma}(\mathbf{i}) = \sum_{\substack{\mathbf{j} \in \{0, 1\}^d \\ \mathbf{i} + \mathbf{j} \in X_{\gamma}(w, d)}} (-1)^{|\mathbf{j}|},
    \label{eqn:weight_vec}
\end{equation}
and the selection region of acceptable product rules defined by
\begin{equation}
    Y_{\gamma}(w, d) = X_{\gamma}(w, d) \backslash X_{\gamma}\bigg(w - \frac{|\gamma|}{\underline{\gamma}}, d\bigg),
    \label{eqn:sel_region}
\end{equation}
where
\begin{equation}
    X_{\gamma}(w, d) = \bigg\{ \mathbf{i} \in \mathbb{N}^{d}_{+}, \mathbf{i} \geq \mathbf{1} : \sum^{d}_{n = 1} (i_n - 1)\gamma_n \leq w \underline \gamma \bigg\},
    \label{eqn:index_set}
\end{equation}
with $\underline \gamma = \min_{1 \leq n \leq d}\gamma_n$ and $|\gamma| = \sum^{d}_{n = 1}\gamma_n$. The \textit{selection criteria} written in (\ref{eqn:sel_region}) and (\ref{eqn:index_set}) can be simplified to the following expression:

\begin{equation}
    w \underline \gamma - |\gamma| < \sum^{d}_{n=1}(i_n - 1)\gamma_n \leq w \underline \gamma.
    \label{eqn:sel_region_simp}
\end{equation}

For a product rule to be included in the sparse grid of sparse grid level $w$, the product rule's level vector $\mathbf{i}$ must satisfy the above criteria.

As stated before, if each dimension is not equally important, then the anisotropic method can account for that by appropriately setting the weight vector. For instance, if the first dimension contributes 25\% more to the variance of the problem than the second, then the ratio of those two components would be $\gamma_2/\gamma_1 = 4/3$. This may seem counterintuitive but in the anisotropic formulation a lower weight value means a higher importance. Additionally, generating the isotropic Smolyak sparse grid is a special case of the anisotropic method. This is done by setting the components of the weight vector to be equal, i.e., $\gamma_1 = \gamma_2 = \cdots = \gamma_d$. In this sense, the anisotropic method is the more generalized version of the sparse grid method.

One more important thing to note about anisotropic sparse grid is that, while the weight vector may be chosen adaptively, as shown in \cite{Gerstner2003}, the relative weight vector can also be chosen using \textit{a priori} or \textit{a posteriori} information instead. In the former, the weight vector can be found by using information about the eigenpairs of the covariance matrix $(\lambda_n, \phi_n)$. In the latter, the dependence of the $n$th input variable can be found by simply ``freezing'' the other input variables to their mean and observe how the output varies \cite{Nobile2008}.

\section{The Coupled KL-Expansion and Anisotropic Sparse Grid applied to the PLF problem}\label{sec:KL+ASG}
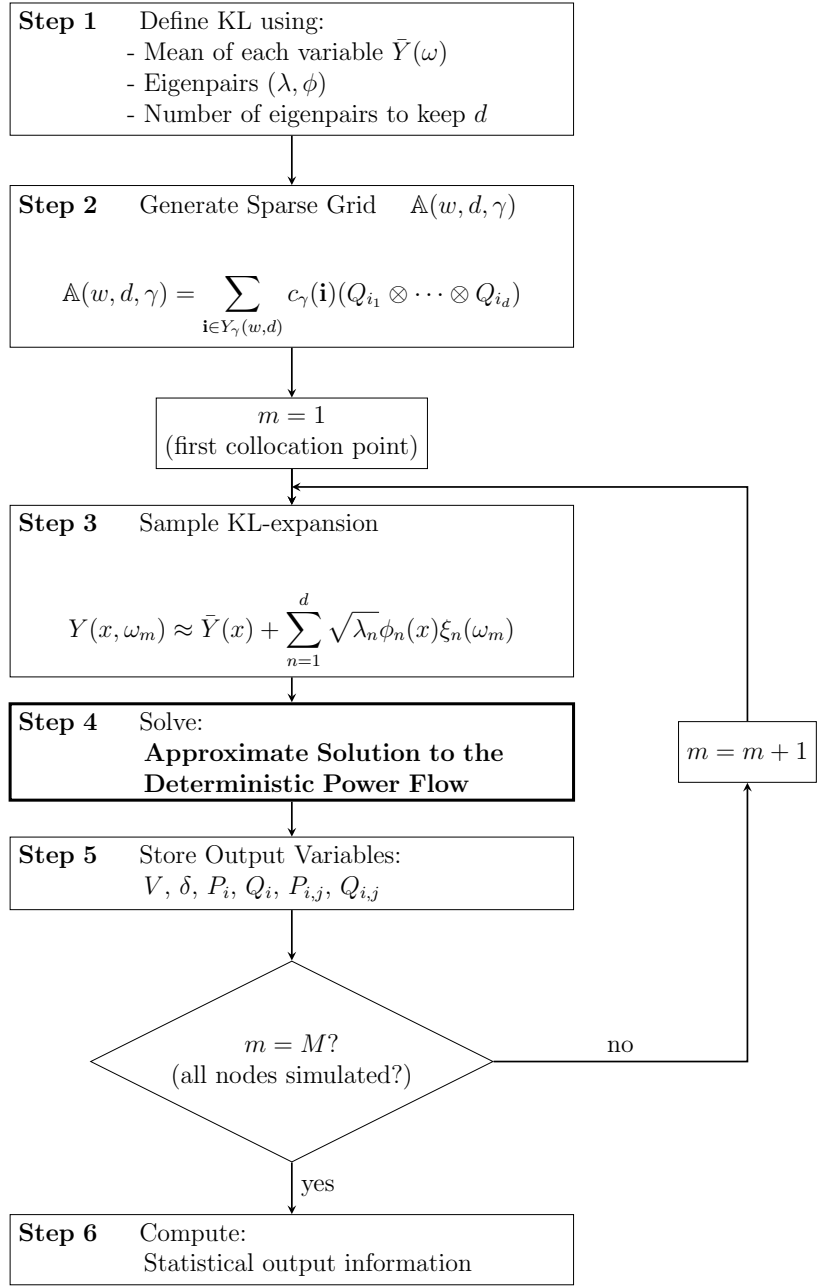
\begin{figure}[!ht]
    \centering
    \resizebox{.8\textwidth}{!}{
    \begin{tikzpicture}[node distance = 2.5cm, auto]
        \node [process, align = left, yshift = 0cm] (pro1) {
        \textbf{Step 1} \quad~Define KL using: \\
        \qquad \qquad ~- Mean of each variable $\Bar{Y}(\omega)$ \\
        \qquad \qquad ~- Eigenpairs $(\lambda, \phi)$ \\
        \qquad \qquad ~- Number of eigenpairs to keep $d$
        };
        \node [process, below of = pro1, align = left, yshift = -.75cm] (pro2) {
        \textbf{Step 2} \quad~Generate Sparse Grid \quad $\mathbb{A}(w, d, \gamma)$ \\
        \begin{equation*}
            \mathbb{A}(w, d, \gamma) = \sum_{\mathbf{i} \in Y_{\gamma}(w, d)} c_{\gamma}(\mathbf{i})(Q_{i_1} \otimes \cdots \otimes Q_{i_d})
        \end{equation*}
        };
        \node[iter, below of = pro2, align = center, yshift = -.25cm] (it1) {$m = 1$ \\ (first collocation point)};
        \node [process, below of = it1, align = left, yshift = -.1cm] (pro3) {
        \textbf{Step 3} \quad~Sample KL-expansion \\
        \begin{equation*}
            Y(x, \omega_m) \approx \Bar{Y}(x) + \sum_{n=1}^{d}\sqrt{\lambda_n}\phi_n(x)\xi_n(\omega_m)
        \end{equation*}
        };
        \node [process, below of = pro3, yshift = -.15cm, align = left, line width = .5mm] (pro5) {
        \textbf{Step 4} \quad~Solve: \\
        \qquad \qquad ~~~\textbf{Approximate Solution to the} \\
        \qquad \qquad ~~~\textbf{Deterministic Power Flow} \\
        };
        \node [process, below of = pro5, align = left, yshift = +.5cm] (pro6) {
        \textbf{Step 5} \quad~Store Output Variables: \\
        \qquad \qquad ~~~$V$, $\delta$, $P_i$, $Q_i$, $P_{i,j}$, $Q_{i,j}$ 
        };
        \node [iter, right of = pro5, xshift = 5cm] (inc) {$m = m + 1$};
        \node [decision, below of = pro6, yshift = -.6cm, align = center] (dec1) {$m = M$?\\(all nodes simulated?)};
        \node [process, below of = dec1, yshift = -.6cm, align = left] (pro7) {
        \textbf{Step 6} \quad~Compute: \\
        \qquad \qquad ~~~Statistical output information};

        \draw [arrow] (pro1) -- (pro2);
        \draw [arrow] (pro2) -- (it1);
        \draw [arrow] (it1) -- (pro3);
        \draw [arrow] (it1) -- coordinate(1) (pro3);
        \draw [arrow] (pro3) -- (pro5);
        \draw [arrow] (pro5) -- (pro6);
        \draw [arrow] (pro6) -- (dec1);
        \draw [arrow] (dec1.south) -- node [pos = 0.5, right] {yes} (pro7);
        \draw [arrow] (dec1.east) -| node [pos = 0.25, above] {no} (inc.south);
        \draw [arrow] (inc.north) |- (1);
    \end{tikzpicture}
    }
    \caption{Flowchart for the proposed algorithm. See Section \ref{sec:KL+ASG} for a joint description of each step of the methodology.}
    \label{fig:PLF_flowchart}
\end{figure}

With the previous sections in mind, Fig.~\ref{fig:PLF_flowchart} depicts the full proposed solution to the PLF problem through the use of a coupled KL-expansion and anisotropic sparse grid method. First, in Step 1 we compute the means of each input variable, the eigenpairs $(\lambda, \phi)$ using (\ref{eqn:frendholm})-(\ref{eqn:frendholm_const}), and determine the $d$ number of eigenpairs to keep. Next, in Step 2 we generate the sparse grid for the stochastic collocation method. As discussed at the beginning of Section~\ref{sec:SC}, it is typical to use cubature rules for the generation of the collocation nodes; the most common of which is the Clenshaw-Curtis quadrature rule \cite{Clenshaw1960}. This is because the Clenshaw-Curtis quadrature rule theoretically produces the fewest number of nodes and will, thereby, be the fastest. Typically, one needs to be cautious in using quadratures rules in high order (dimension) systems. In certain cases, and with some quadrature rules, there can be large negative weights (not to be confused with the anisotropic weight vector) associated with the quadrature nodes. The presence of large negative weights can introduce significant sources of error in the uncertainty approximation. However, it has been shown in \cite{Novak1997} and \cite{Gerstner1998} that, for Smolyak grids, this is not an issue as convergence of the approximation is still guaranteed.

Fej\'{e}r's ``second'' quadrature \cite{Fejer1933} is almost identical to the Clenshaw-Curtis quadrature but with one important difference; the endpoint quadrature nodes are weighted to zero, i.e., $f(-1) = 0$ and $f(+1) = 0$. The Fej\'{e}r second quadrature rule is advantageous because it overcomes a feature of the KL-expansion in which we call ``covariance extrapolation''. Instead of strictly interpolating from the random input variables, the KL-expansion treats the data as samples--in space and random space--of a hypothetical true covariance function, then constructs a low-dimensional representation of this function, and re-samples from it. In this way the KL-expansion produces many more realizations than we originally had. Furthermore, these realizations can potentially have more extreme values than represented in the original data thus giving the appearance of extrapolating the values of the data. By not considering the most extreme samples, we lessen the chance of infeasible input realizations.

After generating the sparse grid in Fig.~\ref{fig:PLF_flowchart}-Step 2 we then assemble the KL-expansion in (\ref{eqn:KL}) and sample the independent random variables $\xi_n$ using the $m$-th collocation point. Subsequently, in Step 4 we solve the deterministic power flow. In Step 5, after each power flow simulation, we store bus voltages ($V$) and angles ($\delta$), bus active ($P_i$) and reactive ($Q_i$) power injections, and branch active ($P_{i,j}$) and reactive ($Q_{i,j}$) flows. After saving all variables the algorithm checks for more collocation points to simulate, if there are, $m$ is incremented and the process is repeated. Otherwise, we proceed to compute relevant statistical information on the output variables to complete the algorithm.

\section{Experimental Setup}\label{sec:setup}
\subsection{Software and Toolboxes for PLF and Sparse Grids}
A MATLAB toolbox set for each of the sparse grids discussed in this paper were originally developed by J.~Burkhardt \cite{Burkhardt}, and we adapted them for use with power systems. In addition to the sparse grid toolboxes, we make use of the open-source power system toolbox MATPOWER \cite{Zimmerman2011} to calculate all deterministic load flow solutions. While there exist other commercially available power system analysis software the sharing of data between these toolboxes reduces computational overhead. All simulations were performed on a machine with Intel Core i7-7700HQ CPU 2.8 GHz PC and 16GB of RAM.

\subsection{Metrics for Performance Comparisons}
In addition to comparing computational burden for our proposed method we use the following equation for statistical accuracy comparisons:
\begin{align}
    \varepsilon_{\mu} = \Bigg|\frac{\mu_{\text{MCS}} - \mu_{\text{grid}}}{\mu_{\text{MCS}}} \Bigg| &&  \varepsilon_{\sigma} = \Bigg|\frac{\sigma_{\text{MCS}} - \sigma_{\text{grid}}}{\sigma_{\text{MCS}}} \Bigg|,
\end{align}
where $\mu_{\text{MCS}}$ and $\sigma_{\text{MCS}}$ represent the mean and standard deviation of the Monte Carlo simulation method and will be henceforth used as reference values. Similarly, $\mu_{\text{grid}}$ and $\sigma_{\text{grid}}$ represent the mean and standard deviation of various sparse grid methods. In addition to error analysis we make use of the Kullback-Leibler divergence (KLD) measure \cite{Kullback1951} to ascertain the likeness of the output distributions of the sparse grid methods to the Monte Carlo method. The KLD measure is defined as follow

\begin{equation}
    \text{KLD}_f = \sum_{x} f_{\text{MCS}}(x)\log\Bigg(\frac{f_{\text{MCS}}(x)}{f_{\text{grid}}(x)}\Bigg),
\end{equation}
where $f_{\text{MCS}}(x)$ and $f_{\text{grid}}(x)$ are the output PDFs of the Monte Carlo method and the various grid methods, respectively. Lastly, the CDFs used for visualization are calculated by creating and sampling the sparse grid interpolant, which approximates (\ref{eqn:inter_operator}), using the method in \cite{Judd2014}.

\section{Experimental Results and Discussion}\label{sec:results}
\subsection{Test Cases for Proposed Methodology}
In this paper, we test the proposed methodology on the IEEE 118-bus system \cite{UWcases} and on a modified version of the IEEE Reliability Test System--1996, the Reliability Test System--Grid Modernization Lab Consortium (RTS-\-GMLC) \cite{Barrows2019}. For the IEEE 118-bus system we will adopt the assumptions first listed in \cite{Morales2007} whereby:
\begin{itemize}
    \item[-] \textbf{Generation units}. Each generation plant is divided into four units with the same power production and a forced outage rate of 0.09. A binomial distribution is used to model each generation plant, and the mean of this input random variable is set to the base case power production of the corresponding generation plant.
    \item[-] \textbf{Load demand}. The active and reactive power of the load buses are modeled as normal distributions, whose means equal the base case data, and whose standard deviations are set arbitrarily as follows: 7\% from bus \#1 to bus \#33, 4\% from bus \#34 to bus \#59, 9\% from bus \#60 to bus \#79, and 5\% from bus \#80 to bus \#118.
\end{itemize}
In total, there are 194 random variables for this test case. Although the random variables considered in the test case are rudimentary, the intent of choosing them is to make comparisons between the classical method presented in \cite{Morales2007} and our proposed method.

The second test case, the RTS-GMLC, is a product of the U.S.~Department of Energy's initiative to modernize the grid. While much of the test case remains the same as the original RTS, e.g.,~transmission line parameters, generator ratings, etc., researchers have mapped the synthetic case over the Southern California, Nevada, and Arizona region. This mapping enables the use of solar and wind resource maps to model the effect of photovoltaic (PV) solar, rooftop photovoltaic (RTPV) solar, concentrated solar power (CSP), wind, and hydro energy sources. A year worth of data for each type of renewable energy is quantized into discrete data points at 1-hour and 5-min intervals. In addition to these generation profiles, the RTS-GMLC dataset includes customer demand which is determined from energy usage data from the various utilities in the region. For this paper we will constrain the problem to using the PV, RTPV, and customer demand information; totaling 59 random variables.

\subsection{Additional Assumptions and Implementation Specifics}
For each test case we choose the number of $d$ dimensions to keep for the truncated KL-expansion such that the percent total contribution of each eigenpair is greater than 90\%. It should be noted that eigenpairs for each random input source are computed separately; that is, the covariance matrix is built using spatial information for each generator type (PV, RTPV, etc.) and load. For example, in the RTS-GMLC test case the covariance matrix for the PV data is a $25\times25$ matrix that explains the joint variability of each PV plant. In the case where the random input source is defined by a PDF and not data, the covariance matrix is built using 10,000 uniformly generated samples of the input source. This assumption leads us to keep 12 (out of 194) and 6 (out of 59) total dimensions for the IEEE 118-bus system and the RTS-GMLC test case, respectively. 

Next, we choose the weight vector $\gamma$ \textit{a priori} using the recursive relationship $\gamma_1 = 1$; $\gamma_n = 2\gamma_{n - 1}$ for $n \leq d$. Similar to the eigenpair calculation, this is done for each type of random input source. This choice means that the largest eigenpairs for each random input source will receive the maximum allowed sampling, denoted as $l_{\text{max}}$, and the proceeding eigenpairs will be less than or equal to $l_{\text{max}}$. For these experiments $l_{\text{max}} = 5$. The Monte Carlo samples used here for performance comparison are taken on the random variables $\xi_n$ in the KL-expansion as compared to randomly sampling the random input variables. We consider representing the input uncertainty as a separate problem and are interested in how sampling the KL-expansion using quadrature nodes compares to random sampling.

Lastly, since we are dealing with numerical methods it is important to mention the convergence and what would cause a power flow to not converge. In this paper we use a simplistic power flow method, which does not guarantee a global, or even local, optimum. Simply put, a simple power flow method tries to balance the generation with demand, how ever it can. Reactive power generator limits are sometimes optionally enforced in order to obtain a first approximation and then refine after reactive power compensation. Convergence (and the rate of convergence) of such methods highly depends on the system's initial conditions. In cases where there are large mismatches between generation and demand the solver may take several hundred iterations to converge, if at all. There does exist globally convergent methods, such as \cite{Ghaddar2016}, which can overcome this problem all together.

\subsection{Accuracy Comparisons}

\begin{figure}[!t]
    \centering
    \includegraphics[trim = {0.0cm 0 0cm 0.75cm}, clip, width = .75\linewidth]{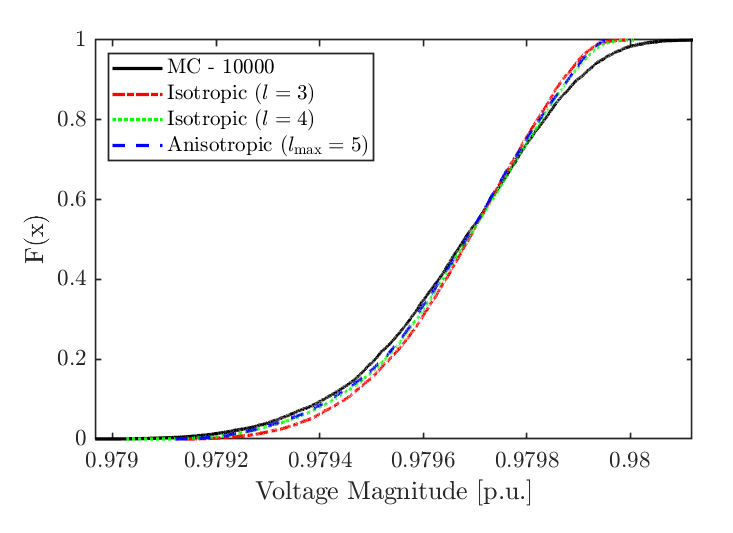}
    \caption{Voltage magnitude at Bus 93 in the IEEE 118-Bus system.}
    \label{fig:volt_118}
\end{figure}
Figure~\ref{fig:volt_118} shows the output voltage at Bus 83 of the IEEE 118-bus system CDF for various levels of isotropic sparse grids, an anisotropic sparse grid with $l_{\text{max}} = 5$, and the reference Monte Carlo simulations which are performed using 10,000 random samples. We observe that the each sparse grid method accurately predicts the response variable with the $l = 3$ isotropic sparse grid performing slightly worse than both the $l = 4$ and the anisotropic sparse grid. This result is expected because as one increases the level of the sparse grid the accuracy of the interpolant also increases.

\begin{table}[!t]\centering
\ra{1.1}
\caption{Statistical accuracy metrics for the IEEE 118-Bus system}
\resizebox{0.75\linewidth}{!}{
\begin{tabular}{cccccccc}
\toprule
\multicolumn{2}{c}{\parbox{0.125\linewidth}{}} & 
\parbox{0.0625\linewidth}{\centering $V$} & 
\parbox{0.0625\linewidth}{\centering $\delta$} & 
\parbox{0.0625\linewidth}{\centering $P_i$} & 
\parbox{0.0625\linewidth}{\centering $Q_i$} &
\parbox{0.0625\linewidth}{\centering $P_{i,j}$} & 
\parbox{0.0625\linewidth}{\centering $Q_{i,j}$} \\
\midrule
\multirow{3}{2cm}{\centering Isotropic ($l = 3$)} & $\varepsilon_{\mu}$ [\%] & 0.01 & 0.05 & 0.03 & 0.09 & 0.11 & 0.04 \\
& $\varepsilon_{\sigma}$ [\%] & 4.16 & 0.71 & 12.0 & 5.22 & 6.78 & 5.28 \\
& $\text{KLD}_f$ & 0.84 & 0.05 & 0.06 & 0.12 & 0.18 & 0.13\\
\midrule
\multirow{3}{2cm}{\centering Isotropic ($l = 4$)} & $\varepsilon_{\mu}$ [\%] & 0.01 & 0.05 & 0.03 & 0.09 & 0.11 & 0.04 \\
& $\varepsilon_{\sigma}$ [\%] & 4.16 & 0.71 & 12.0 & 5.22 & 6.78 & 5.28 \\
& $\text{KLD}_f$ & 0.56 & 0.01 & 0.04 & 0.08 & 0.13 & 0.09 \\
\midrule
\multirow{3}{2.5cm}{\centering Anisotropic ($l_\text{max} = 5$)} & $\varepsilon_{\mu}$ [\%] & 0.01 & 0.05 & 0.03 & 0.09 & 0.11 & 0.04 \\
& $\varepsilon_{\sigma}$ [\%] & 4.70 & 0.68 & 13.0 & 6.00 & 7.71 & 5.90 \\
& $\text{KLD}_f$ & 0.63 & 0.02 & 0.04 & 0.08 & 0.14 & 0.09 \\
\bottomrule
\end{tabular}
}
\label{tbl:error_118}
\end{table}

Looking at the accuracy metrics listed in Table~\ref{tbl:error_118} we see that there is no difference between the two isotropic sparse grids. We suspect this is the result of the relatively low variance of the input sources. With respect to the KLD measure we should keep in mind that lower numbers indicate a better approximation of the true solution. The ideal approximation is one where the KLD measure is, in fact, zero. Here, by comparing the KLD measure between the two isotropic sparse grids we notice that we do indeed produce a more accurate statistical approximation with a higher level sparse grid. However, the big takeaway from this experiment, which will be further supported by the next section, is that the anisotropic sparse grid is able to maintain a comparable accuracy with far fewer collocation points. In this experiment, there is only 0.5-1\% difference in accuracy.

\begin{figure}[!t]
    \centering
    \subfloat
    {\centering
        {
        \includegraphics[viewport = 12 350 475 700, clip, width = 0.485\linewidth]{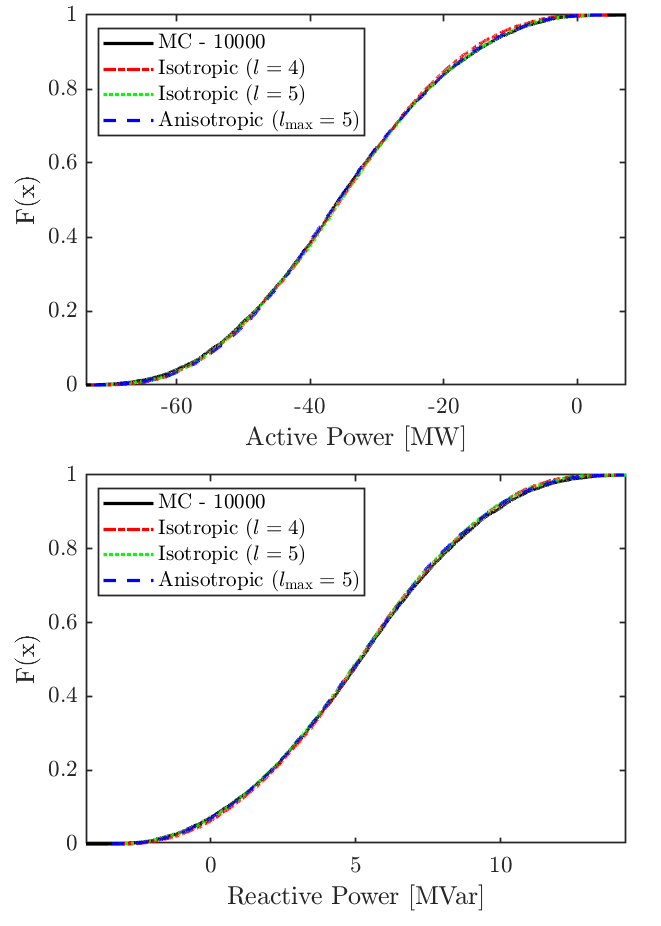}
        }
    }\hspace{-6pt}%
    \subfloat
    {\centering
        {
        \includegraphics[viewport = 12 0 475 350, clip, width = 0.485\linewidth]{Figures/GMLC_res2.png}
        }
    }
    \caption{Reliability Test System--Grid Modernization Lab Consortium (RTS-GMLC) active and reactive power line flows for the branch between Bus 203 and Bus 209. Negative values indicate the direction of the flow, therefore, in this case active power is flowing predominantly from Bus 209 to Bus 203.}
    \label{fig:branch_GMLC}
\end{figure}

\begin{table}[!t]\centering
\ra{1.1}
\caption{Statistical accuracy metrics for RTS-GMLC}
\resizebox{0.75\linewidth}{!}{
\begin{tabular}{cccccccc}
\toprule
\multicolumn{2}{c}{\parbox{0.125\linewidth}{}} & 
\parbox{0.0625\linewidth}{\centering $V$} & 
\parbox{0.0625\linewidth}{\centering $\delta$} & 
\parbox{0.0625\linewidth}{\centering $P_i$} & 
\parbox{0.0625\linewidth}{\centering $Q_i$} &
\parbox{0.0625\linewidth}{\centering $P_{i,j}$} & 
\parbox{0.0625\linewidth}{\centering $Q_{i,j}$} \\
\midrule
\multirow{3}{2cm}{\centering Isotropic ($l = 4$)} & $\varepsilon_{\mu}$ [\%] & 1e-3 & 0.40 & 0.21 & 0.69 & 0.86 & 0.45 \\
& $\varepsilon_{\sigma}$ [\%] & 1.20 & 0.27 & 3.21 & 1.03 & 1.48 & 1.84 \\
& $\text{KLD}_f$ & 0.38 & 5e-3 & 0.03 & 3e-3 & 8e-3 & 0.01 \\
\midrule
\multirow{3}{2cm}{\centering Isotropic ($l = 5$)} & $\varepsilon_{\mu}$ [\%] & 1e-3 & 0.40 & 0.21 & 0.69 & 0.86 & 0.45 \\
& $\varepsilon_{\sigma}$ [\%] & 1.20 & 0.27 & 3.21 & 1.03 & 1.48 & 1.84 \\
& $\text{KLD}_f$ & 0.19 & 3e-3 & 0.02 & 3e-3 & 6e-3 & 7e-3 \\
\midrule
\multirow{3}{2.5cm}{\centering Anisotropic ($l_\text{max} = 5$)} & $\varepsilon_{\mu}$ [\%] & 1e-3 & 0.40 & 0.21 & 0.69 & 0.86 & 0.45 \\
& $\varepsilon_{\sigma}$ [\%] & 1.20 & 0.27 & 3.21 & 1.06 & 1.48 & 1.91 \\
& $\text{KLD}_f$ & 0.44 & 3e-3 & 0.02 & 3e-3 & 6e-3 & 7e-3 \\
\bottomrule
\end{tabular}
}
\label{tbl:error_GMLC}
\end{table}

Figure~\ref{fig:branch_GMLC} shows the CDF for active and reactive power line flow from Bus 203 to Bus 209 of the RTS-GMLC test case. As we can see, each of the three sparse grid methods closely match the Monte Carlo simulations used as a reference. The differences between the two methods are even harder to see in Table~\ref{tbl:error_GMLC}. Theory tells us that as we increase the level of sparse grid, i.e., the number of collocation points, we should expect to see a higher degree of accuracy. This is not evident by looking the $\varepsilon_{\mu}$ and $\varepsilon_{\sigma}$ measures alone. Upon inspecting the KLD measure does one see that the $l = 5$ is better than the $l = 4$ sparse grid. More importantly, the KLD measure shows us that the proposed anisotropic sparse grid method competes with the $l = 5$ isotropic sparse method while having roughly 2,000 fewer collocation points.

\subsection{Computational Burden Comparisons}
Computational burden comparisons for the IEEE 118-bus system are found in Table~\ref{tbl:compu_118}. To see which portion of the method takes the longest to perform we have broken up the total time spent into four segments:~eigenpair calculation, sparse grid construction, KL-expansion, and deterministic load flow solutions. Again, for comparison we will use 10,000 Monte Carlo simulations which, in this case, take 34.3 s to complete. The first thing to notice is the total number of samples--or collocation points for the sparse grids--required per method. We observe that as the number of points, i.e., level, increases, the total computational burden increases as well. This is a factor of it both taking longer to generate the sparse grid points (as shown in $t_{\text{grid}}$) as well as the time it takes to perform each deterministic load flow. By comparing each method we see that the anisotropic sparse grid interpolant is roughly 1.7x faster than the next fastest sparse grid method. As we have mentioned earlier, this significant decrease in computational time only costs 0.5 - 1\% decrease in accuracy. We see a similar result for the RTS-GMLC test case in Table~\ref{tbl:compu_GMLC}. In this case, the anistropic sparse grid is only 0.05 seconds faster than the $l = 4$ isotropic sparse grid but, again, provides the same level of accuracy as the $l = 5$ isotropic sparse grid.

\begin{table}[t]\centering
\ra{1.1}
\caption{Computational comparison for the anisotropic sparse grid method for the PLF problem on the IEEE 118-bus system}

\resizebox{.8\linewidth}{!}{
\begin{threeparttable}
\begin{tabular}{ccccc}\toprule
\parbox{0\linewidth}{}& 
\parbox{0.1\linewidth}{\centering MC-10000} & 
\parbox{0.2\linewidth}{\centering Isotropic Sparse Grid ($l = 3$)} & 
\parbox{0.2\linewidth}{\centering Isotropic Sparse Grid ($l = 4$)} & 
\parbox{0.2\linewidth}{\centering Anisotropic Sparse Grid ($l_{\text{max}} = 5$)} \\
\midrule
$N_{\text{samples}}$\tnote{\dag}    &   10,000  &   337      &   3,249    &   213  \\
$t_{\text{eigenpairs}}$ &   0.01 s  &   0.01 s   &   0.01 s   &   0.01 s \\
$t_{\text{grid}}$       &   -       &   0.08 s   &   0.75 s   &   0.19 s \\
$t_{\text{KL}}$         &   1.10 s  &   0.01 s   &   0.04 s   &   0.01 s \\
$t_{\text{P.F.}}$       &   34.3 s  &   1.50 s   &   11.8 s   &   0.88 s \\
\midrule
\textbf{Total time}     &   \textbf{35.4 s}    &   \textbf{1.60 s}      &   \textbf{12.6 s}     &   \textbf{1.09 s}  \\
\bottomrule
\end{tabular}
\begin{tablenotes}\footnotesize
    \item[\dag] Out of the original 194 random variables 12 are kept after dimension reduction
\end{tablenotes}
\end{threeparttable}
}

\label{tbl:compu_118}
\end{table}

\begin{table}[!t]\centering
\ra{1.1}
\caption{Computational comparison for the anisotropic sparse grid method for the PLF problem on the RTS-GMLC system}
\resizebox{.8\linewidth}{!}{
\begin{threeparttable}

\begin{tabular}{ccccc}\toprule
\parbox{0\linewidth}{}& 
\parbox{0.1\linewidth}{\centering MC-10000} & 
\parbox{0.2\linewidth}{\centering Isotropic Sparse Grid ($l = 4$)} & 
\parbox{0.2\linewidth}{\centering Isotropic Sparse Grid ($l = 5$)} & 
\parbox{0.2\linewidth}{\centering Anisotropic Sparse Grid ($l_{\text{max}} = 5$)} \\
\midrule
$N_{\text{samples}}$\tnote{\dag}    &   10,000  &   545      &   2,561    &   489  \\
$t_{\text{eigenpairs}}$ &   0.03 s  &   0.03 s   &   0.03 s   &   0.03 s \\
$t_{\text{grid}}$       &   -       &   0.10 s   &   0.57 s   &   0.15 s \\
$t_{\text{KL}}$         &   0.56 s  &   0.01 s   &   0.03 s   &   0.02 s \\
$t_{\text{P.F.}}$       &   30.3 s  &   1.92 s   &   8.15 s   &   1.81 s \\
\midrule
\textbf{Total time}     &   \textbf{30.9 s}    &   \textbf{2.06 s}      &   \textbf{8.78 s}     &   \textbf{2.01 s}  \\
\bottomrule
\end{tabular}
\begin{tablenotes}\footnotesize
    \item[\dag] Out of the original 59 random variables 6 are kept after dimension reduction
\end{tablenotes}
\end{threeparttable}
}
\label{tbl:compu_GMLC}
\end{table}

\begin{figure}[!t]
    \centering
    \includegraphics[trim = {0.0cm 0 0cm 0.75cm}, clip, width = .7\linewidth]{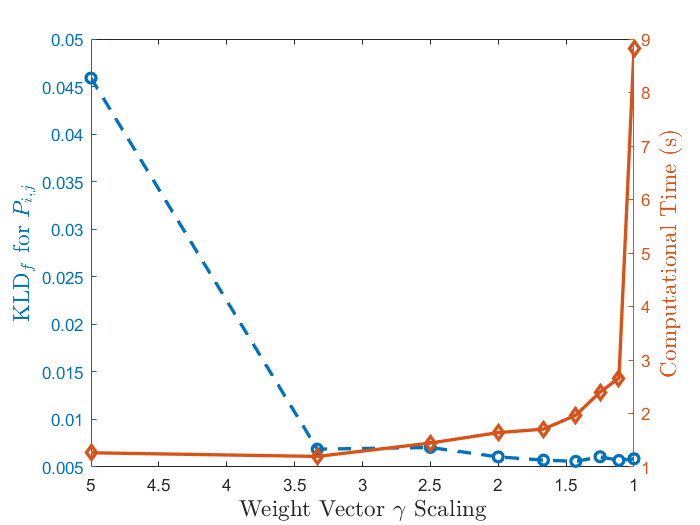}
    \caption{Comparison between computational time and accuracy, both as a function of the anisotropic weight vector $\gamma$. Graphs are generated using the RTS-GMLC test case and setup as described in Section~6.1.}
    \label{fig:KLD_v_Time}
\end{figure}

In Figure~\ref{fig:KLD_v_Time}, we show how the proposed method scales, both in accuracy and computational time, as a function of dimension importance. In Section 6.2 we establish that the weight vector used in these simulations is calculated using a recursive relationship. To generate the plots in Fig.~\ref{fig:KLD_v_Time} we modify the equation by adding a scalar $\zeta$ such that $\gamma_n = \zeta\gamma_{n-1}$ for $n \leq d$. In this equation, when $\zeta$ is large the dimensions after the first are treated with less importance. When $\zeta = 1$ each dimension is treated with the same importance and we create an fully isotropic sparse grid. This plot shows that as we move towards an isotropic sparse grid (increased collocation points) the estimator becomes more accurate. The more points simulated in each dimension, the better the uncertainty can be represented. This, however, comes at an exponential increase in the computation time. This plot goes to show that not only are the number of dimensions important to consider but also the number of collocation points as well.
    
\subsection{Methods to Increase Performance}
One very useful feature of the proposed method is that there are several ways to tune the method for either increases in accuracy and/or decreases in computational burden. The first way is to change the number of dimensions to keep in the KL-expansion. Reducing the number of eigenpairs to keep will significantly decrease the number of total collocation points need to be created, and thereby decreasing computational burden. In most cases, the first eigenpair $\{\lambda_1, \phi_1\}$ accounts for greater than 80\% of the total variance in the system which leads us to question how important the proceeding eigenpairs are to the accuracy of the system. The second way is parallelizing the deterministic load flow solutions. Depending on the number of cores available and the access to, for example, the Parallel Computing Toolbox\textsuperscript{TM} in MATLAB, parallelizing this segment of the code requires changing one line of code. Third, increasing or decreasing $l_{\text{max}}$ changes the maximum allowed level for the anisotropic sparse grid. Increasing $l_{\text{max}}$ means having more collocation nodes for each dimension, i.e., a higher level of accuracy, but at the cost of increased computational burden. The opposite is true for a decrease in $l_{\text{max}}$. Lastly, changing the weight vector $\gamma$ will have a similar effect as changing $l_{\text{max}}$. One way to increase the performance is to weigh higher dimension by their normalized eigenvalues. This will have the outcome to increase/decrease the number of KL-expansion terms to keep and will impact the accuracy and/or computational burden.

\section{Conclusion}\label{sec:conclusion}
In this paper, we have established a formal methodology for implementing the coupled KL-expansion and anisotropic sparse grid to solve the PLF problem. Using two test cases, the IEEE 118-bus system and the more modernized grid RTS-GMLC, we have shown that our methodology is able to alleviate the ``curse of dimensionality'' issue present in the PLF problem. Furthermore, our method can solve the high-dimensional load flow problem 30 times faster while only having a 0.05\% and 6\% error in the mean and standard deviation when compared to 10,000 Monte Carlo simulations. In the modernized RTS-GMLC we show that, by taking advantage of the correlated input data, we are able to accurately and efficiently approximate their effect on the output variables. In addition to the observed high performance, we discuss how to tune the method to further increase accuracy and/or decrease computational burden, the easiest of which to implement is the parallelization of the deterministic load flow problem. Depending on the infrastructure available, this change could enable this method to be used as a real-time assessment of uncertainty given a large number of random variables.

\bibliographystyle{els/elsarticle-num-names}
\bibliography{mendeley.bib}

\begin{thebibliography}{38}
\expandafter\ifx\csname natexlab\endcsname\relax\def\natexlab#1{#1}\fi
\providecommand{\url}[1]{\texttt{#1}}
\providecommand{\href}[2]{#2}
\providecommand{\path}[1]{#1}
\providecommand{\DOIprefix}{doi:}
\providecommand{\ArXivprefix}{arXiv:}
\providecommand{\URLprefix}{URL: }
\providecommand{\Pubmedprefix}{pmid:}
\providecommand{\doi}[1]{\href{http://dx.doi.org/#1}{\path{#1}}}
\providecommand{\Pubmed}[1]{\href{pmid:#1}{\path{#1}}}
\providecommand{\bibinfo}[2]{#2}
\ifx\xfnm\relax \def\xfnm[#1]{\unskip,\space#1}\fi
\bibitem[{Borkowska(1974)}]{Borkowska1974}
\bibinfo{author}{B.~Borkowska},
\newblock \bibinfo{title}{{Probabilistic Load Flow}},
\newblock \bibinfo{journal}{IEEE Trans. Power Appar. Syst.}
  \bibinfo{volume}{PAS-93} (\bibinfo{year}{1974}) \bibinfo{pages}{1--6}.
  \DOIprefix\doi{10.1109/TPAS.1974.293973}.
\bibitem[{Morales and P{\'{e}}rez-Ruiz(2007)}]{Morales2007}
\bibinfo{author}{J.~M. Morales}, \bibinfo{author}{J.~P{\'{e}}rez-Ruiz},
\newblock \bibinfo{title}{{Point estimate schemes to solve the probabilistic
  power flow}},
\newblock \bibinfo{journal}{IEEE Trans. Power Syst.} \bibinfo{volume}{22}
  (\bibinfo{year}{2007}) \bibinfo{pages}{1594--1601}.
  \DOIprefix\doi{10.1109/TPWRS.2007.907515}.
\bibitem[{Chen et~al.(2008)Chen, Chen, and Bak-Jensen}]{Chen2008}
\bibinfo{author}{P.~Chen}, \bibinfo{author}{Z.~Chen},
  \bibinfo{author}{B.~Bak-Jensen},
\newblock \bibinfo{title}{{Probabilistic load flow: A review}},
\newblock \bibinfo{journal}{3rd Int. Conf. Deregul. Restruct. Power Technol.
  DRPT 2008}  (\bibinfo{year}{2008}) \bibinfo{pages}{1586--1591}.
  \DOIprefix\doi{10.1109/DRPT.2008.4523658}.
\bibitem[{Allan et~al.(1981)Allan, da~Silva, and Burchett}]{Allan1981}
\bibinfo{author}{R.~N. Allan}, \bibinfo{author}{A.~M.~L. da~Silva},
  \bibinfo{author}{R.~C. Burchett},
\newblock \bibinfo{title}{{Evaluation methods and accuracy in probabilistic
  load flow solutions}},
\newblock \bibinfo{journal}{IEEE Trans. Power Appar. Syst.}
  \bibinfo{volume}{PAS-100} (\bibinfo{year}{1981}) \bibinfo{pages}{2539--2546}.
  \DOIprefix\doi{10.1109/TPAS.1981.316721}.
\bibitem[{Kloubert and Rehtanz(2017)}]{Kloubert2017}
\bibinfo{author}{M.~L. Kloubert}, \bibinfo{author}{C.~Rehtanz},
\newblock \bibinfo{title}{{Enhancement to the combination of point estimate
  method and Gram-Charlier Expansion method for probabilistic load flow
  computations}},
\newblock \bibinfo{journal}{IEEE Manchester PowerTech}  (\bibinfo{year}{2017}).
  \DOIprefix\doi{10.1109/PTC.2017.7980901}.
\bibitem[{Zhang and Lee(2004)}]{Zhang2004}
\bibinfo{author}{P.~Zhang}, \bibinfo{author}{S.~T. Lee},
\newblock \bibinfo{title}{{Probabilistic Load Flow Computation Using the Method
  of Combined Cumulants and Gram-Charlier Expansion}},
\newblock \bibinfo{journal}{IEEE Trans. Power Syst.} \bibinfo{volume}{19}
  (\bibinfo{year}{2004}) \bibinfo{pages}{676--682}.
  \DOIprefix\doi{10.1109/TPWRS.2003.818743}.
\bibitem[{Fan et~al.(2012)Fan, Member, Vittal, Heydt, and Fellow}]{Fan2012}
\bibinfo{author}{M.~Fan}, \bibinfo{author}{S.~Member},
  \bibinfo{author}{V.~Vittal}, \bibinfo{author}{G.~T. Heydt},
  \bibinfo{author}{L.~Fellow},
\newblock \bibinfo{title}{{Probabilistic Power Flow Studies for Transmission
  Systems With Photovoltaic Generation Using Cumulants}},
\newblock \bibinfo{journal}{IEEE Trans. Power Syst.} \bibinfo{volume}{27}
  (\bibinfo{year}{2012}) \bibinfo{pages}{2251--2261}.
  \DOIprefix\doi{10.1109/TPWRS.2012.2190533}.
\bibitem[{Appino et~al.(2017)Appino, M{\"{u}}hlpfordt, Faulwasser, and
  Hagenmeyer}]{Appino2017}
\bibinfo{author}{R.~R. Appino}, \bibinfo{author}{T.~M{\"{u}}hlpfordt},
  \bibinfo{author}{T.~Faulwasser}, \bibinfo{author}{V.~Hagenmeyer},
\newblock \bibinfo{title}{{On Solving Probabilistic Load Flow for Radial Grids
  using Polynomial Chaos}},
\newblock in: \bibinfo{booktitle}{2017 IEEE Manchester PowerTech},
  \bibinfo{year}{2017}, pp. \bibinfo{pages}{1--6}.
\bibitem[{Wu et~al.(2017)Wu, Zhou, Dong, and Song}]{Wu2017}
\bibinfo{author}{H.~Wu}, \bibinfo{author}{Y.~Zhou}, \bibinfo{author}{S.~Dong},
  \bibinfo{author}{Y.~Song},
\newblock \bibinfo{title}{{Probabilistic Load Flow Based on Generalized
  Polynomial Chaos}},
\newblock \bibinfo{journal}{IEEE Trans. Power Syst.} \bibinfo{volume}{32}
  (\bibinfo{year}{2017}) \bibinfo{pages}{820--821}.
\bibitem[{Sun et~al.(2018)Sun, Tu, Chen, Zhang, and Duan}]{Sun2018}
\bibinfo{author}{X.~Sun}, \bibinfo{author}{Q.~Tu}, \bibinfo{author}{J.~Chen},
  \bibinfo{author}{C.~Zhang}, \bibinfo{author}{X.~Duan},
\newblock \bibinfo{title}{{Probabilistic load flow calculation based on sparse
  polynomial chaos expansion}},
\newblock \bibinfo{journal}{IET Gener. Transm. Distrib.} \bibinfo{volume}{12}
  (\bibinfo{year}{2018}) \bibinfo{pages}{2735--2744}.
  \DOIprefix\doi{10.1049/iet-gtd.2017.0859}.
\bibitem[{Zhou et~al.(2017)Zhou, Wu, Gu, and Song}]{Zhou2017}
\bibinfo{author}{Y.~Zhou}, \bibinfo{author}{H.~Wu}, \bibinfo{author}{C.~Gu},
  \bibinfo{author}{Y.~Song},
\newblock \bibinfo{title}{{A Novel Method of Polynomial Approximation for
  Parametric Problems in Power Systems}},
\newblock \bibinfo{journal}{IEEE Trans. Power Syst.} \bibinfo{volume}{32}
  (\bibinfo{year}{2017}) \bibinfo{pages}{3298--3307}.
  \DOIprefix\doi{10.1109/TPWRS.2016.2623820}.
\bibitem[{Rosenblueth(1981)}]{Rosenblueth1981}
\bibinfo{author}{E.~Rosenblueth},
\newblock \bibinfo{title}{{Two-point estimates in probabilities}},
\newblock \bibinfo{journal}{Appl. Math. Model.} \bibinfo{volume}{5}
  (\bibinfo{year}{1981}) \bibinfo{pages}{329--335}.
  \DOIprefix\doi{10.1016/S0307-904X(81)80054-6}.
\bibitem[{Harr(1989)}]{Harr1989}
\bibinfo{author}{M.~E. Harr},
\newblock \bibinfo{title}{{Probabilistic estimates for multivariate analyses}},
\newblock \bibinfo{journal}{Appl. Math. Model.} \bibinfo{volume}{13}
  (\bibinfo{year}{1989}) \bibinfo{pages}{313--318}.
  \DOIprefix\doi{10.1016/0307-904X(89)90075-9}.
\bibitem[{Li(1992)}]{Li1992}
\bibinfo{author}{K.~S. Li},
\newblock \bibinfo{title}{{Point Estimate Method for Calculating Statistical
  Moments}},
\newblock \bibinfo{journal}{J. Eng. Mech.} \bibinfo{volume}{118}
  (\bibinfo{year}{1992}) \bibinfo{pages}{1506--1511}.
  \DOIprefix\doi{10.1061/(ASCE)0733-9399(1992)118:7(1506)}.
\bibitem[{Hong(1998)}]{Hong1998}
\bibinfo{author}{H.~Hong},
\newblock \bibinfo{title}{{An efficient point estimate method for probabilistic
  analysis}},
\newblock \bibinfo{journal}{Reliab. Eng. Syst. Saf.} \bibinfo{volume}{59}
  (\bibinfo{year}{1998}) \bibinfo{pages}{261--267}.
  \DOIprefix\doi{10.1016/S0951-8320(97)00071-9}.
\bibitem[{Xiu(2009)}]{Xiu2009}
\bibinfo{author}{D.~Xiu},
\newblock \bibinfo{title}{{Fast numerical methods for stochastic computations:
  A review}},
\newblock in: \bibinfo{booktitle}{Commun. Comput. Phys.}, \bibinfo{year}{2009},
  pp. \bibinfo{pages}{242--272}. \DOIprefix\doi{10.1.1.148.5499}.
  \href{http://arxiv.org/abs/arXiv:1011.1669v3}{{\tt arXiv:arXiv:1011.1669v3}}.
\bibitem[{Xiu(2010)}]{Dongbin2010}
\bibinfo{author}{D.~Xiu}, \bibinfo{title}{{Numerical Methods for Stochastic
  Computations : A Spectral Method Approach}}, \bibinfo{publisher}{Princeton
  University Press}, \bibinfo{year}{2010}.
\bibitem[{Smolyak(1963)}]{Smolyak1963}
\bibinfo{author}{S.~A. Smolyak},
\newblock \bibinfo{title}{Quadrature and interpolation formulas for tensor
  products of certain classes of functions},
\newblock \bibinfo{journal}{Doklady Akademii Nauk} \bibinfo{volume}{148}
  (\bibinfo{year}{1963}) \bibinfo{pages}{1042--1045}.
\bibitem[{Tang et~al.(2016)Tang, Ni, Ponci, and Monti}]{Tang2016}
\bibinfo{author}{J.~Tang}, \bibinfo{author}{F.~Ni}, \bibinfo{author}{F.~Ponci},
  \bibinfo{author}{A.~Monti},
\newblock \bibinfo{title}{{Dimension-Adaptive Sparse Grid Interpolation for
  Uncertainty Quantification in Modern Power Systems: Probabilistic Power
  Flow}},
\newblock \bibinfo{journal}{IEEE Trans. Power Syst.} \bibinfo{volume}{31}
  (\bibinfo{year}{2016}) \bibinfo{pages}{907--919}.
  \DOIprefix\doi{10.1109/TPWRS.2015.2404841}.
\bibitem[{{Ni} et~al.(2017){Ni}, {Nguyen}, and {Cobben}}]{Ni2017}
\bibinfo{author}{F.~{Ni}}, \bibinfo{author}{P.~H. {Nguyen}},
  \bibinfo{author}{J.~F.~G. {Cobben}},
\newblock \bibinfo{title}{Basis-adaptive sparse polynomial chaos expansion for
  probabilistic power flow},
\newblock \bibinfo{journal}{IEEE Transactions on Power Systems}
  \bibinfo{volume}{32} (\bibinfo{year}{2017}) \bibinfo{pages}{694--704}.
  \DOIprefix\doi{10.1109/TPWRS.2016.2558622}.
\bibitem[{Kosambi(1943)}]{Kosambi1943}
\bibinfo{author}{D.~D. Kosambi},
\newblock \bibinfo{title}{Statistics in function space},
\newblock \bibinfo{journal}{The Journal of the Indian Mathematical Society}
  \bibinfo{volume}{7} (\bibinfo{year}{1943}) \bibinfo{pages}{76--88}.
\bibitem[{Karhunen(1947)}]{Karhunen1947}
\bibinfo{author}{K.~Karhunen},
\newblock \bibinfo{title}{{On linear methods in probability and statistics}},
\newblock \bibinfo{journal}{Ann. Acad. Sci. Fenn. Ser. AI Math.-Phys}
  (\bibinfo{year}{1947}) \bibinfo{pages}{1----79}.
\bibitem[{Lo{\`{e}}ve(1978)}]{Loeve1978}
\bibinfo{author}{M.~Lo{\`{e}}ve}, \bibinfo{title}{{Probability Theory II}},
  volume~\bibinfo{volume}{46} of \textit{\bibinfo{series}{Graduate Texts in
  Mathematics}}, \bibinfo{publisher}{Springer New York}, \bibinfo{address}{New
  York, NY}, \bibinfo{year}{1978}. \DOIprefix\doi{10.1007/978-1-4684-9464-8}.
\bibitem[{Nobile et~al.(2008)Nobile, Tempone, and Webster}]{Nobile2008}
\bibinfo{author}{F.~Nobile}, \bibinfo{author}{R.~Tempone},
  \bibinfo{author}{C.~G. Webster},
\newblock \bibinfo{title}{{An Anisotropic Sparse Grid Stochastic Collocation
  Method for Partial Differential Equations with Random Input Data}},
\newblock \bibinfo{journal}{SIAM J. Numer. Anal.} \bibinfo{volume}{46}
  (\bibinfo{year}{2008}) \bibinfo{pages}{2411--2442}.
  \DOIprefix\doi{10.1137/070680540}. \href{http://arxiv.org/abs/1404.2647}{{\tt
  arXiv:1404.2647}}.
\bibitem[{Griebel and Grestner(1998)}]{Griebel1998a}
\bibinfo{author}{M.~Griebel}, \bibinfo{author}{T.~Grestner},
\newblock \bibinfo{title}{{Numerical Integration using Sparse-grids}},
\newblock \bibinfo{journal}{Numer. Algorithms} \bibinfo{volume}{18}
  (\bibinfo{year}{1998}) \bibinfo{pages}{209--232}.
\bibitem[{Gerstner and Griebel(2003)}]{Gerstner2003}
\bibinfo{author}{T.~Gerstner}, \bibinfo{author}{M.~Griebel},
\newblock \bibinfo{title}{{Dimension – Adaptive Tensor – Product
  Quadrature}},
\newblock \bibinfo{journal}{Computing} \bibinfo{volume}{71}
  (\bibinfo{year}{2003}) \bibinfo{pages}{65--87}.
  \DOIprefix\doi{10.1007/s00607-003-0015-5}.
\bibitem[{Burkardt(2012)}]{Burkardt2012}
\bibinfo{author}{J.~Burkardt},
\newblock \bibinfo{title}{{The ``Combining Coefficient'' for Anisotropic Sparse
  Grids}},
\newblock \bibinfo{journal}{Interdiscip. Cent. Appl. Math. \& Inf. Technol.
  Dep. Virginia Tech}  (\bibinfo{year}{2012}) \bibinfo{pages}{1--14}.
  \URLprefix
  \url{https://people.sc.fsu.edu/~jburkardt/presentations/sgmga_coefficient.pdf}.
\bibitem[{Clenshaw and Curtis(1960)}]{Clenshaw1960}
\bibinfo{author}{C.~W. Clenshaw}, \bibinfo{author}{A.~R. Curtis},
\newblock \bibinfo{title}{A method for numerical integration on an automatic
  computer},
\newblock \bibinfo{journal}{Numerische Mathematik} \bibinfo{volume}{2}
  (\bibinfo{year}{1960}) \bibinfo{pages}{197--205}.
\bibitem[{Novak and Ritter(1997)}]{Novak1997}
\bibinfo{author}{E.~Novak}, \bibinfo{author}{K.~Ritter},
\newblock \bibinfo{title}{The curse of dimension and a universal method for
  numerical integration},
\newblock in: \bibinfo{booktitle}{Multivariate approximation and splines},
  \bibinfo{publisher}{Springer}, \bibinfo{year}{1997}, pp.
  \bibinfo{pages}{177--187}.
\bibitem[{Gerstner and Griebel(1998)}]{Gerstner1998}
\bibinfo{author}{T.~Gerstner}, \bibinfo{author}{M.~Griebel},
\newblock \bibinfo{title}{Numerical integration using sparse grids},
\newblock \bibinfo{journal}{Numerical algorithms} \bibinfo{volume}{18}
  (\bibinfo{year}{1998}) \bibinfo{pages}{209}.
\bibitem[{Fej\'{e}r(1933)}]{Fejer1933}
\bibinfo{author}{L.~Fej\'{e}r},
\newblock \bibinfo{title}{Mechanische quadraturen mit positiven cotesschen
  zahlen},
\newblock \bibinfo{journal}{Mathematische Zeitschrift} \bibinfo{volume}{37}
  (\bibinfo{year}{1933}) \bibinfo{pages}{287--309}.
\bibitem[{Burkardt(2019)}]{Burkhardt}
\bibinfo{author}{J.~Burkardt}, \bibinfo{title}{{John Burkhardt's Home Page}},
  \bibinfo{year}{2019}. \URLprefix \url{https://people.sc.fsu.edu/~jburkardt/}.
\bibitem[{Zimmerman et~al.(2011)Zimmerman, Murilla-Sanchez, and
  Thomas}]{Zimmerman2011}
\bibinfo{author}{R.~D. Zimmerman}, \bibinfo{author}{C.~E. Murilla-Sanchez},
  \bibinfo{author}{R.~J. Thomas},
\newblock \bibinfo{title}{{MATPOWER: Steady-State Operations, Planning and
  Analysis Tools for Power Systems Research and Education}},
\newblock \bibinfo{journal}{IEEE Trans. Power Syst.} \bibinfo{volume}{26}
  (\bibinfo{year}{2011}) \bibinfo{pages}{12--19}.
\bibitem[{Kullback and Leibler(1951)}]{Kullback1951}
\bibinfo{author}{S.~Kullback}, \bibinfo{author}{R.~A. Leibler},
\newblock \bibinfo{title}{On information and sufficiency},
\newblock \bibinfo{journal}{The annals of mathematical statistics}
  \bibinfo{volume}{22} (\bibinfo{year}{1951}) \bibinfo{pages}{79--86}.
\bibitem[{Judd et~al.(2014)Judd, Maliar, Maliar, and Valero}]{Judd2014}
\bibinfo{author}{K.~L. Judd}, \bibinfo{author}{L.~Maliar},
  \bibinfo{author}{S.~Maliar}, \bibinfo{author}{R.~Valero},
\newblock \bibinfo{title}{{Smolyak method for solving dynamic economic models:
  Lagrange interpolation, anisotropic grid and adaptive domain}},
\newblock \bibinfo{journal}{J. Econ. Dyn. Control} \bibinfo{volume}{44}
  (\bibinfo{year}{2014}) \bibinfo{pages}{92--123}.
  \DOIprefix\doi{10.1016/j.jedc.2014.03.003}.
\bibitem[{Christie(2000)}]{UWcases}
\bibinfo{author}{R.~Christie}, \bibinfo{title}{{Power Systems Test Case
  Archive}}, \bibinfo{year}{2000}. \URLprefix
  \url{http://labs.ece.uw.edu/pstca/}.
\bibitem[{{Barrows} et~al.(2019){Barrows}, {Bloom}, {Ehlen}, {Ikaheimo},
  {Jorgenson}, {Krishnamurthy}, {Lau}, {McBennett}, {O'Connell}, {Preston},
  {Staid}, {Stephen}, and {Watson}}]{Barrows2019}
\bibinfo{author}{C.~{Barrows}}, \bibinfo{author}{A.~{Bloom}},
  \bibinfo{author}{A.~{Ehlen}}, \bibinfo{author}{J.~{Ikaheimo}},
  \bibinfo{author}{J.~{Jorgenson}}, \bibinfo{author}{D.~{Krishnamurthy}},
  \bibinfo{author}{J.~{Lau}}, \bibinfo{author}{B.~{McBennett}},
  \bibinfo{author}{M.~{O'Connell}}, \bibinfo{author}{E.~{Preston}},
  \bibinfo{author}{A.~{Staid}}, \bibinfo{author}{G.~{Stephen}},
  \bibinfo{author}{J.~{Watson}},
\newblock \bibinfo{title}{The ieee reliability test system: A proposed 2019
  update},
\newblock \bibinfo{journal}{IEEE Transactions on Power Systems}
  (\bibinfo{year}{2019}) \bibinfo{pages}{1--1}.
  \DOIprefix\doi{10.1109/TPWRS.2019.2925557}.
\bibitem[{{Ghaddar} et~al.(2016){Ghaddar}, {Marecek}, and
  {Mevissen}}]{Ghaddar2016}
\bibinfo{author}{B.~{Ghaddar}}, \bibinfo{author}{J.~{Marecek}},
  \bibinfo{author}{M.~{Mevissen}},
\newblock \bibinfo{title}{Optimal power flow as a polynomial optimization
  problem},
\newblock \bibinfo{journal}{IEEE Transactions on Power Systems}
  \bibinfo{volume}{31} (\bibinfo{year}{2016}) \bibinfo{pages}{539--546}.

\end{thebibliography}

\end{document}